\documentstyle[aps,12pt,cite]{revtex}

\begin{document}

\title {Quantum-mechanical tunneling: differential operators, zeta-functions and
determinants } 

\author { J. Casahorr\'an \footnote{email:casahorran@eupla.unizar.es}}
\address{Departamento de F\'{\i}sica Te\'orica \footnote{DFTUZ/00/09}, \\
Universidad de Zaragoza, E-50009 Zaragoza, Spain}

\maketitle

\begin{abstract}

We consider in detail the quantum-mechanical problem associated with the
motion of a one-dimensional particle under the action of the double-well
potential. Our main tool will be the euclidean (imaginary time) version
of the path-integral method. Once we perform the 
the Wick rotation, the euclidean equation
of motion is the same as the usual one for the point particle in real time, except
that the potential at issue is turned upside down. In doing so, our double-well
potential becomes a two-humped potential. As required by the semiclassical
approximation we may study the quadratic fluctuations over the instanton which
represents in this context the localised finite-action solutions of the
euclidean equation of motion. The determinants of the quadratic differential operators
are evaluated by means of the zeta-function method. We write in closed form the 
eigenfunctions as well as the energy eigenvalues  corresponding to such operators 
by using the shape-invariance symmetry. The effect of the multi-instantons
configurations is also included in this approach.
 
\end{abstract}

\vfill \eject

\section {Introduction.}

The semiclassical approximation to the euclidean path-integral represents one
of the most valuable tools in quantum theory. By expanding around topological
configurations one can obtain information about the tunneling phenomenon in
physical systems ranging from non-relativistic one-particle quantum mechanics
to relativistic gauge field theories. The basis of this approach relies on the
so-called instanton which represents a localised finite-action solution of the
euclidean equation of motion where the time variable is essentially imaginary.
To sum up, we first find the appropiate classical configuration and subsequently
evaluate the quadratic fluctuations. The functional integration is solved in
terms of the gaussian scheme except for the zero-modes which appear by virtue
of the translational invariance of the system. It is customarily assumed the
existence of  collective coordinates so that as a matter of fact
one performs the gaussian integration in the directions orthogonal to the
zero-modes. \par

To illustrate the usual instanton method in a simple model we resort to the
double-well potential as first considered by Polyakov \cite{po} in his
pioneering work on the subject. The famous problem of the level splittings
is analyzed in terms of the instanton calculus which represents by itself
an alternative to the WKB technique. From a physical point of view the
result  can be summarized as follows: the quantum mechanical
amplitude consists on an exponential of the classical action of the topological
configuration multiplied by a fluctuation factor which takes into account
the inverse square root of the determinant of harmonic eigenmodes. In this
paper we will not try to expose new fundamental developments on the matter
but rather focus the attention to the explicit evaluation of the determinants
at issue. \par

By analogy with the finite-dimensional case one can suspect that a functional
determinant includes an infinite product of eigenvalues. As expected
this expression will be highly divergent but in any case we may regularize it
by means of the ratio of two determinants. When the eigenfunctions as well as
the energy eigenvalues of the quadratic operators are known in closed form
the zeta-function method  serves to evaluate this quantity in a
systematic way \cite{ha}. For the double-well potential it proves convenient
to take advantage of the shape-invariance symmetry so that the only tools
we need are supersymmetric quantum mechanics and the physics of the
free-particle. \par

In order to obtain an accurate description of the level splitting  for
a double-well potential (or double-humped once the potential is turned
upside down) we introduce a chain of instantons and anti-instantons. If
the distance between the centers of such configurations is large the
interaction can be neglected. This framework represents the dilute gas
approximation in quantum theory. The article is organised as follows. First
of all we expose in detail the instantonic approach in non-relativistic
one-dimensional particle mechanics. To illustrate the procedure in a simple
but relevant context we consider the aforementioned double-well potential.
We keep for the last section the analysis of the effect of the multi-instantons
configurations. \par  

\section {Instantons in particle mechanics.}

In this section we describe the instanton calculus for the one-dimensional
spinless particle as can be found for instance in \cite{kl}. For
reference we assume that our particle moves under the effect of a confining
potential $V(x)$ so that the quantum model exhibits a pure discrete spectrum
of energy eigenvalues. Whitout loss of generality we choose the origin of
the energy such that the minimum (minima) of the potential satisfy $V(x) = 0$.
Now taking the mass of the particle equal to unity, i.e. $m = 1$, the
lagrangian which governs the behaviour of the model corresponds to

\begin{equation}
L = {{1} \over {2}} \left( {{dx} \over {dt}} \right)^2 - V(x)
\label{eq:1}
\end{equation}

If the particle is located at the initial time $t_i = -T/2$ at the point
$x_i$ while one finds it when $t_f = T/2$ at the point $x_f$, the
functional version of the non-relativistic quantum mechanics allows us to
express the transition amplitude in terms of a sum over all paths joining
the world points with coordinates $(-T/2, x_i)$ and $(T/2, x_f)$. The main
physical ingredient consists in the way in which we weight the trajectories
according to the imaginary exponential of the action. At this point it proves convenient
to write the action $S$ starting from the lagrangian $L$, i.e.

\begin{equation}
S = \int_{-T/2}^{T/2} L(x,\dot{x}) \ dt
\label{eq:2}
\end{equation}

\noindent so that the transition amplitude is

\begin{equation}
<x_f\vert \exp(-i H T) \vert x_i> = N(T) \int [dx] \exp i S[x(t)]
\label{eq:3}
\end{equation}

It may be worth spelling out that $H$ represents the standard hamiltonian
of the model while the symbol $[dx]$ serves to indicate the integration over all
functions which fulfill the aforementioned boundary conditions. The factor
$N(T)$ will be adjusted later on to make the whole integral finite and suitable
normalised. As the hamiltonian $H$ yields a pure discrete spectrum of energy
eigenvalues, namely

\begin{equation}
H \vert n > = E_n \vert n >
\label{eq:4}
\end{equation}

\noindent we have that 

\begin{equation}
<x_f\vert \exp(-i H T) \vert x_i> = \sum_{n} \exp(-i E_n T) <x_f \vert n >
<n\vert x_i >
\label{eq:5}
\end{equation}

As this last expression is difficult to handle due to the oscillating
character of the exponential, we perform the transition to the imaginary
time. Making the substitution $t \rightarrow - i \tau$, known in the
literature as the Wick rotation, one can take advantage of the euclidean
formalism where

\begin{equation}
i S[x(t)] \rightarrow \int_{-T/2}^{T/2} \left[ - {{1} \over {2}} 
\left( {{dx} \over {d\tau}} \right)^2 - V(x) \right]  \ d\tau
\label{eq:6}
\end{equation}

Now it remains to identify the so-called euclidean action $S_e$, i.e.

\begin{equation}
S_e =  \int_{-T/2}^{T/2} \left[  {{1} \over {2}} 
\left( {{dx} \over {d\tau}} \right)^2 + V(x) \right]  \ d\tau
\label{eq:7}
\end{equation}

The euclidean version of (\ref{eq:3}) corresponds to

\begin{equation}
<x_f\vert \exp(- H T) \vert x_i> = N(T) \int [dx]  \ \exp - S_e[x(\tau)]
\label{eq:8}
\end{equation}

In order to explain the meaning of the semiclassical approximation we consider
a function $x_c(\tau)$ satisfying the boundary conditions at issue. Now we
expand a general $x(\tau)$ with the same boundary conditions according to

\begin{equation}
x(\tau) = x_c(\tau) + \sum_j c_j \  x_j(\tau)
\label{eq:9}
\end{equation}

\noindent where $x_j(\tau)$ constitute a complete set of orthonormal functions

\begin{equation}
\int_{-T/2}^{T/2} x_j(\tau) \  x_k(\tau) \ d\tau = \delta_{jk}
\label{eq:10}
\end{equation}

\noindent  vanishing at the boundary

\begin{equation}
x_j(\pm T/2) = 0
\label{eq:11}
\end{equation}

In addition the $[dx]$ can be chosen as

\begin{equation}
[dx] = \prod_j {{dc_j} \over {\sqrt{2 \pi}}}
\label{eq:12}
\end{equation}

The semiclassical approximation takes for granted that $x_c(\tau)$ represents
a stationary point of the euclidean action, thus satisfying the equation

\begin{equation}
{{d^2x} \over {d\tau^2}} = V^{\prime}(x)
\label{eq:13}
\end{equation}

\noindent where the prime denotes as usual the derivative with respect to the spatial
coordinate. We recognize in (\ref{eq:13}) the euclidean equation of motion
for our particle once the potential has been turned upside down. Now the
crucial feature is the analysis of the second variation of the action with
respect to $x_c(\tau)$. (There is no linear variation due to the stationary
point character of $x_c(\tau)$). In doing so we find that

\begin{equation}
S_e[x_c(\tau) + \delta x(\tau)] = S_{eo} + 
\int_{-T/2}^{T/2} \delta x \left[- {{ 1} \over {2}} {{d^2} \over {d\tau^2}} \delta x +
{{ 1} \over {2}} V^{\prime \prime}[x_c(\tau)] \delta x \right] \ d\tau
\label{eq:14}
\end{equation}

\noindent being $S_{eo}$ the classical action associated with the configuration
$x_c(\tau)$. The procedure becomes much more accesible in terms of a complete
set of eigenfunctions (eigenvalues) of the so-called stability equation

\begin{equation}
-  {{d^2} \over {d\tau^2}} v_j(\tau) +
 V^{\prime \prime}[x_c(\tau)] v_j(\tau) = \epsilon_j v_j(\tau)
\label{eq:15}
\end{equation}

\noindent so that the family just anticipated in (\ref{eq:10}) is obtained
by means of (\ref{eq:15}). The problem becomes diagonal in this new scheme
so that

\begin{equation}
S_e = S_{eo} + {{ 1} \over {2}} \sum_j c_j^2 \epsilon_j 
\label{eq:16}
\end{equation}

Taking into account the standard gaussian integral, i.e.

\begin{equation}
\int_{-\infty}^{\infty} \exp(- \lambda z^2 /2) \ dz = \sqrt{{{2\pi} \over {\lambda}}}
\ \ \ \ ( \lambda > 0) 
\label{eq:17}
\end{equation}

\noindent the euclidean transition amplitude reduces to

\begin{equation}
<x_f\vert \exp(- H T) \vert x_i> = N(T) \exp(- S_{eo}) \ \prod_j 
\epsilon_{j}^{-1/2}
\label{eq:18}
\end{equation}

Almost from the very beginning of the subject the product of eigenvalues
has been written as 
 
\begin{equation}
\prod_j \epsilon_{j}^{-1/2} = \left\{ Det \left[
-  {{d^2} \over {d\tau^2}}  +
 V^{\prime \prime}[x_c(\tau)]  \right] \right\}^{-1/2}
\label{eq:19}
\end{equation}

\noindent a notation which obviously comes from the finite-dimensional case.
It should be emphasized that by now we assume that all the eigenvalues
$\epsilon_j$ are positive to avoid problems of definition for (\ref{eq:19})
due to the existence of zero-modes (more on this later). The physical
meaning of (\ref{eq:18}) can be summarized as follows: the transition
amplitude consists on an exponential of the classical action of the
configuration $x_c(\tau)$ at issue multiplied by a factor which takes into
account the inverse square root of the determinant of a second order 
differential operator. \par

To close this section we focus the interest on the explicit evaluation of the
euclidean transition amplitude for the well-grounded problem of the harmonic
oscillator. To be precise we consider boundary conditions such that
$x_i = x_f = 0$. The purpose of the calculation is twofold. First of all we 
find physical arguments to fix the factor $N(T)$. On the other hand the
harmonic oscillator acts as reference to deal with the ratio of determinants
when we go to more complicated systems. Let us explore therefore the case where

\begin{equation}
V(x) = {{\nu^2 } \over {2}} \  x^2
\label{eq:20}
\end{equation}

\noindent so that $V^{\prime \prime} (x = 0 ) = \nu^2$. If the potential is turned
upside down the euclidean equation of motion dictates that the only dynamical
possibility starting from $x_i = 0$ at $t_i = - T/2$ is $x_c(\tau) = 0$ whenever
we require also that $x_f = 0$ when $t_f =  T/2$. Otherwise the particle scapes
towards plus or minus infinity. As the classical action associated with this trivial
path vanishes the semiclassical approximation reduces itself to

\begin{equation}
<x_f = 0\vert \exp(- H_{ho} T) \vert x_i = 0> = N(T) \left\{ Det \left[
-  {{d^2} \over {d\tau^2}}  +
 \nu^2  \right]\right\}^{-1/2}
\label{eq:21}
\end{equation}

For the explicit evaluation of (\ref{eq:21}) we follow the pattern exposed in
\cite{ra}. To start from scratch we write a complete set of eigenfunctions
of the quadratic operator whose determinant we desire to compute. To be precise

\begin{equation}
x_j = \cos \lambda \tau
\label{eq:22}
\end{equation}

\begin{equation}
\tilde{x}_j = \sin \lambda \tau
\label{eq:23}
\end{equation}

Once we add the boundary conditions (see (\ref{eq:11})) the eigenvalues of
energy $\epsilon_j$ which correspond to (\ref{eq:22}) and (\ref{eq:23})
are of the form

\begin{equation}
\epsilon_j = {{j^2 \pi^2} \over {T^2}} + \nu^2 \ \ \ \ \  (j = 1, 2, ...)
\label{eq:24}
\end{equation}

Now the principal tool for extracting the physical information hidden in the
product of eigenvalues is the factorization of the determinant, namely

\begin{equation}
N(T) \ \left\{ Det \left[
-  {{d^2} \over {d\tau^2}} + \nu^2  \right]\right\}^{-1/2} = N(T)
 \left[\prod_{j=1}^{\infty} {{j^2 \pi^2} \over {T^2}} 
 \right]^{-1/2}
\left[\prod_{j=1}^{\infty} \left(1 + {{\nu^2 T^2} \over {j^2 \pi^2}} \right)
 \right]^{-1/2}
\label{eq:25}
\end{equation}

As can be seen easily the two first factors represent the euclidean transition
amplitude for the free-particle so that

\begin{equation}
N(T) \left[\prod_{j=1}^{\infty} {{j^2 \pi^2} \over {T^2}} 
\right]^{-1/2} = <x_f = 0\vert \exp(- H_o T) \vert x_i = 0>
\label{eq:26}
\end{equation}

\noindent where $H_o$ corresponds to

\begin{equation}
H_o = - {{d^2} \over {d\tau^2}}
\label{eq:27}
\end{equation}

Now it suffices to resort to a set of plane waves labeled by $\vert k >$ to
obtain that

\begin{equation}
N(T) \left[\prod_{j=1}^{\infty} {{j^2 \pi^2} \over {T^2}} 
\right]^{-1/2} = \int_{-\infty}^{\infty} \exp(- k^2 T/2) \ {{dk} \over {2\pi}}
\label{eq:28}
\end{equation}

Using again the formula (\ref{eq:17}) we find 

\begin{equation}
N(T) \left[\prod_{j=1}^{\infty} {{j^2 \pi^2} \over {T^2}} 
\right]^{-1/2} = {{1} \over {\sqrt{2 \pi T}}}
\label{eq:29}
\end{equation}

The last factor of (\ref{eq:25}) can be transformed into a more suitable
expression according to \cite{gr}

\begin{equation}
\pi z \prod_{j=1}^{\infty} \left(1 + {{z^2} \over {j^2}}\right) =
\sinh \pi z
\label{eq:30}
\end{equation}

\noindent where in our case it happens that

\begin{equation}
z = {{\nu T} \over {\pi}}
\label{eq:31}
\end{equation}

With all these partial results at hand the final form for the amplitude written in
(\ref{eq:21}) should be 

\begin{equation}
<x_f = 0\vert \exp(- H_{ho} T) \vert x_i = 0> = \left({{\nu} \over {\pi}}\right)^{1/2} \
\left(2 \sinh \nu T \right)^{-1/2}
\label{eq:32}
\end{equation}

For reference we point out the behaviour of (\ref{eq:32}) when considering the
limit $T \rightarrow \infty$ since this final step is crucial for instance
in the analysis of the double-well potential to get valuable information
about the splitting. To sum up

\begin{equation}
<x_f = 0\vert \exp(- H_{ho} T) \vert x_i = 0> \  \longrightarrow \ 
 \left({{\nu} \over {\pi}}\right)^{1/2} \
\exp(- \nu T/2)  
\left[1 + {{1} \over {2}} \exp(- 2 \nu T) + ... \  \right]
\label{eq:33}
\end{equation}

Although it is a simple matter to check in (\ref{eq:32}) how this semiclassical approximation is exact
for the harmonic oscillator we prefer to go to more relevant physical systems where
the euclidean version of the path-integral represents probably the most elegant
tool to study the tunneling phenomenon. \par

\section {The double-well potential.}

Once we have understood the main properties of the instanton calculus for the
one-dimensional particle we proceed to examine the method in a concrete
example. For such a purpose one needs a particle moving in a potential well
which has a set of degenerate minima. No matter what the specific details of the
potential, it is the case that the euclidean equation of motion exhibits 
topologically nontrivial solutions with finite-action. As anticipated in the
previous section these solutions connect adjoining minima of the potential at issue
and make the transition through a region classically forbidden. 
Althoug a periodic potential should be an excellent benchmark in this context
we consider now the double-well potential $V(x)$ given by

\begin{equation}
V(x) = {{\omega^2 } \over {8}} (x^2 - 1)^2
\label{eq:34}
\end{equation}

From a classical point of
view there are two minima located at $x_{-} = - 1$ and $x_{+} =  1$ as
expected when dealing with a potential which enjoys the discrete symmetry
$x \rightarrow - x$. One could just as well assume that $\omega^2 \gg 1$ so
that the barrier is high enough to decompose the system into a sum of two
independent harmonic oscillators widely separated from each other. Then the
lowest state of the model as a whole has a twofold degeneracy. Qualitatively
speaking the particle lives in the right well or in the  left one
while executes small oscillations around the points $x_{-} = - 1$ or $x_{+} =  1$.
For a finite barrier however, the particle in either of the two wells has a 
non-vanishing transition amplitude to tunnel into the other well so that the
individual wave functions for left-hand and right-hand harmonic oscillators
will mix. To sum up, the discrete symmetry $x \rightarrow - x$ is not
spontaneously broken since the expectation value of the coordinate $x$
evaluated for the ground state is zero as expected for a wave function which
makes an even superposition of wave functions centered at either of the two
left and right wells. \par

Now we can carry things further and discuss how the tunneling phenomenon
emerges in the euclidean framework. The analysis is based on the existence
of a topological configuration which connects the two minima located at
$x =\pm 1$. The solution of the euclidean equation of motion which makes
the connection is precisely the instanton according to the name first
coined in the mid-seventies. Following the conventional wisdom the
instanton interpolates between adjoining minima of the potential and lifts
the degeneracy of the classical vacua. The instanton only appears once
the potential is turned upside down. In doing so the two-well model transforms
into a double-humped potential. At first glance it seems
physically relevant to take into account the transition between the points
$x_{-} = - 1$ and $x_{+} =  1$ whenever we have previously managed to obtain the
topological solution in closed form. \par

Therefore we are interested in a trajectory which starts at $t_i = - T/2$ from
the point $x_{i} = - 1$ and reaches $x_{f} =  1$ when $t_f =  T/2$. It is
customarily assumed that $T \rightarrow \infty$ since the solution of the
problem for finite $T$ is much more complicated (more on this later). This
movement, where the particle itself takes an infinite time to climb up 
(or drop from) the small piece next to the top of the two potential mountains,
can be understood in terms of the conservation law of the euclidean energy once
the double-well is just reversed. Simply put, we solve the problem by
integration of a first-order differential equation instead of resorting
to the conventional euclidean equation of motion written in (\ref{eq:13}). As
the euclidean energy is equal to zero for the motion of our particle, the
form of the topological solution is quite simple, namely

\begin{equation}
x_c(\tau) =  \tanh {{\omega (\tau - \tau_c) } \over {2}} 
\label{eq:35}
\end{equation}

\noindent where the parameter $\tau_c$ indicates the point at which the instanton
makes the jump.  
Of course the so-called antiinstanton is obtained by means of the
transformation $\tau \rightarrow - \tau$ and allows the connection between
$x_{i} =  1$ and $x_{f} = - 1$. Now we can compute the action associated with
$x_c(\tau)$ according to (\ref{eq:7}) so that

\begin{equation}
S_{eo} = {{2 \omega} \over {3}} 
\label{eq:36}
\end{equation}

It may be interesting at this point to remind that actually we need configurations
for which $x$ is equal to $\pm 1$ at large but finite values $\tau = \pm T/2$.
However the topological solution written in (\ref{eq:35}) reaches the points
$\pm 1$ at infinite times. Fortunately the difference is so small that it can
be ignored since at the end of the procedure we are interested precisely in the
large $T$ limit. As required by the semiclassical approximation we must calculate
the quadratic fluctuations correction once the classical contribution is known.
The question we wish to address now is the analysis of the transition amplitude
between $x_{i} = - 1$ and $x_{f} =  1$ taking as reference the harmonic oscillator
derived from

\begin{equation}
V^{\prime\prime} ( x = \pm 1) = \omega^2
\label{eq:37}
\end{equation}

Our description takes over

\begin{eqnarray*}
<x_f = 1\vert \exp(- H T) \vert x_i = - 1> = N(T) \left\{ Det \left[
- {{d^2} \over {d\tau^2}}  +
 \omega^2  \right]\right\}^{-1/2}   
\end{eqnarray*}
\begin{equation}
\left\{{{Det \left[- (d^2/d\tau^2) + V^{\prime \prime}[x_c(\tau)] \right]} \over
{Det \left[- (d^2/d\tau^2) + \omega^2 \right]}}\right\}^{- 1/2} \ 
\exp(-S_{eo})
\label{eq:38}
\end{equation}
 
\noindent where we have multiplied and divided by the determinant for the
harmonic oscillator precisely. Incorporating the explicit form of $x_c(\tau)$, where
we fix by now $\tau_c = 0$, the stability equation reads

\begin{equation}
-  {{d^2} \over {d\tau^2}} x_j(\tau) +
 \left[\omega^2 - {{3\omega^2} \over {2 \cosh^2 (\omega \tau/2)}}\right]
 x_j(\tau) = \epsilon_j x_j(\tau)
\label{eq:39}
\end{equation}

\noindent and corresponds to a Schrodinger's equation with Posch-Teller potential.
Although the pro\-blem can be studied in more conventional forms we prefer to
do that by means of the shape-invariance symmetry \cite{ju}. Ultimately the
only tools we need are supersymmetric quantum mechanics and the free particle.
(The details can be consulted in Appendix A). Among other things one finds a
zero-mode $x_o(\tau)$ which in principle could jeopardize the evaluation of the
determinant. However this eigenvalue $\epsilon_o = 0$ comes by no surprise
since it reflects the translational invariance of the system. In other words,
there is one direction in the functional space of the second variations which
is incapable of changing the action. As a matter of fact one can discover the
existence of a zero-mode starting from (\ref{eq:13}). It suffices an additional
derivative with respect to $\tau$ to find that

\begin{equation}
x_{o}(\tau) = {{1} \over {\sqrt{S_{eo}}}} {{dx_{c}} \over {d\tau}}
\label{eq:40}
\end{equation}

\noindent is just the solution of (\ref{eq:15}) with $\epsilon_o = 0$. Notice
the normalization of $x_{o}(\tau)$ which is due precisely to the zero 
euclidean energy condition for $x_{c}(\tau)$. The way out of this apparent cul-de-sac
is simple. The integration over $c_{o}$ (see (\ref{eq:9})) becomes equivalent
to the integration over the center of the instanton $\tau_{c}$. To fix the
jacobian of the transformation involved we take a first change such that

\begin{equation}
\Delta x(\tau) = x_{o}(\tau) \  \Delta c_{o}
\label{eq:41}
\end{equation}

According to the general expression written in (\ref{eq:9}) we find that under
a shift $\Delta \tau_{c}$ the effect should be

\begin{equation}
\Delta x(\tau) = - \sqrt{S_{eo}} \ x_{o}(\tau) \Delta \tau_{c}
\label{eq:42}
\end{equation}

Now the identification between (\ref{eq:41}) and (\ref{eq:42}) yields

\begin{equation}
d c_{o} = \sqrt{S_{eo}} d\tau_{c}
\label{eq:43}
\end{equation}

\noindent where the minus sign disappears since what matters is the modulus
of the jacobian at issue. In doing so we have that

\begin{eqnarray*}
\left\{{{Det \left[- (d^2/d\tau^2) + V^{\prime \prime}[x_c(\tau)] \right]} \over
{Det \left[- (d^2/d\tau^2) + \omega^2 \right]}}\right\}^{- 1/2} = 
 \end{eqnarray*}
\begin{equation}
\left\{{{Det^{\prime} \left[- (d^2/d\tau^2) + V^{\prime \prime}[x_c(\tau)] \right]} \over
{Det \left[- (d^2/d\tau^2) + \omega^2 \right]}}\right\}^{- 1/2} \
\sqrt{{{S_{eo}} \over {2 \pi}}} \ d\tau_{c}
\label{eq:44}
\end{equation}

\noindent where $Det^{\prime}$ stands for the so-called reduced determinant
once the zero-mode has been removed. \par

Different methods can be found in the literature to find the ratio of
determinants written in (\ref{eq:44}). One takes firstly the system enclosed
in a box of lenght $T$, thus avoiding the subleties associated with the
continuous spectrum \cite{kl}. At the end of the whole procedure
 it suffices the limit $T \rightarrow \infty$ to achieve a
meaningful result. 
However we prefer from the very beginning to work in open space and evaluate
the ratio of determinants by using the zeta-function method. (The interested
reader can find in the Appendix B a sketch of this technique as tailored to
our needs). Going back to (\ref{eq:39}) and (\ref{eq:44}), it suffices to
make the change of variable

\begin{equation}
z = {{\omega \tau} \over {2}}
\label{eq:45}
\end{equation}

\noindent to recognize the presence of the ratio of determinants (see Appendix B)

\begin{equation}
Q_{2} = {{Det^{\prime} \ O_{2}} \over {Det \ P_{2}}}
\label{eq:46}
\end{equation}

\noindent togheter with a global factor $\beta$ given by

\begin{equation}
\beta = {{\omega^2} \over {4}}
\label{eq:47}
\end{equation}

As regards the non-zero spectrum of the operator $O_{2}$ we have a discrete
level with $E_{1} = 3$ while the energy of the scattering states corresponds to

\begin{equation}
E_{k} = k^2 + 4
\label{eq:48}
\end{equation}

\noindent for eigenfunctions derived from the plane waves according to

\begin{equation}
\phi_{2,k}(z) = {{A_{2}^{\dagger}(z)} \over {\sqrt{k^2 + 4}}} \
{{A_{1}^{\dagger}(z)} \over {\sqrt{k^2 + 1}}} \ \left[{{\exp(ikz)} \over
{\sqrt{2 \pi}}} \right]
\label{eq:49}
\end{equation}

The explicit form of the operators $A_{2}^{\dagger}(z)$ and
$A_{1}^{\dagger}(z)$, namely

\begin{equation}
A_{1}^{\dagger} = - {{d} \over {dz}} +  \tanh z
\label{eq:50}
\end{equation}

\begin{equation}
A_{2}^{\dagger} = - {{d} \over {dz}} + 2 \tanh z
\label{eq:51}
\end{equation}

\noindent allows us to find that

\begin{equation}
\phi_{2,k}(z) = \left(- k^2 + 2 - {{3} \over {\cosh^2 z}} - 3 i k \tanh z\right)
\ {{\exp(ikz)} \over {\sqrt{2 \pi} \  \sqrt{k^2 + 4} \  \sqrt{k^2 + 1}}}
\label{eq:52}
\end{equation}

As a natural consequence of (\ref{eq:52}) the regularized spectral density
$\rho_{r}(k)$ of the problem reads
 
\begin{equation}
\rho_{r}(k) = - {{3 (k^2 + 2)} \over {\pi (k^2 + 4) (k^2 + 1)}}
\label{eq:53}
\end{equation}
 
In this scheme the suitable zeta-function $\zeta_r(s)$ to evaluate the
ratio of determinants of (\ref{eq:44}) can be understood as

\begin{equation}
\zeta_r(s) = \zeta_{O_{2}}(s) - \zeta_{P_{2}}(s)
\label{eq:54}
\end{equation}

This information is sufficient to write

\begin{equation}
\zeta_r(s) = {{1} \over {\Gamma(s)}} \int_0^{\infty} \mu^{s-1}  d\mu  
\left[\exp(- 3 \mu)  - {{3} \over {\pi}}
\int_{-\infty}^{\infty}  {{ (k^2 + 2) \exp[-(k^2 + 4) \mu] } \over
 { (k^2 + 4) (k^2 + 1)}} \  dk \right]
\label{eq:55}
\end{equation}

\noindent which, in turn, gives rise to \cite{gr}

\begin{equation}
\zeta_r(s) = {{1} \over {3^{s}}} - {{3} \over {\pi}}
\int_{-\infty}^{\infty}  {{ (k^2 + 2) } \over
 {  (k^2 + 1) (k^2 + 4)^{s+1}}} \  dk 
\label{eq:56}
\end{equation}

Breaking the integral of (\ref{eq:56}) into more simple components it is
not difficult to obtain $\zeta_r(s)$ in terms of Gamma and Hypergeometric
Functions, namely \cite{gr}

\begin{equation}
\zeta_r(s) = {{1} \over {3^{s}}} - {{3} \over {\sqrt{\pi}}} 
{{1} \over {2^{2s+1}}} {{\Gamma \left(s + {{1} \over {2}}\right)} \over {\Gamma(s + 1)}}
- {{3} \over {\sqrt{\pi}}} 
{{1} \over {2^{2s+3}}} {{\Gamma \left(s + {{3} \over {2}}\right)} \over {\Gamma(s + 2)}}
 \ F\left(1, s + {{3} \over {2}}, s + 2, {{3} \over {4}}\right)
\label{eq:57}
\end{equation}

Taking into account that \cite{ab}

\begin{equation}
F\left(1, {{3} \over {2}}, 2, {{3} \over {4}}\right) = {{8} \over {3}}
\label{eq:58}
\end{equation}

\noindent the evaluation of $\zeta_r(s)$ at $s = 0$ yields

\begin{equation}
\zeta_r(0) = - 1
\label{eq:59}
\end{equation}

Let us come down to the concrete details which allow the evaluation of 
$\zeta_r^{\prime}(0)$. Our approach provides a result of the form

\begin{equation}
\zeta_r^{\prime}(0) = - \ln 3 + 8 \ln 2 - {{1} \over {2}} - {{3} \over {16}}
F^{\prime}(s + {{3} \over {2}},1,s + 2, {{3} \over {4}})\vert_{s=0} 
\label{eq:60}
\end{equation}

\noindent once one incorporates that  \cite{ab}

\begin{equation}
\Gamma^{\prime}({{1} \over {2}}) = - \sqrt{\pi} \  (\gamma + 2 \ln 2) 
\label{eq:61}
\end{equation}

\begin{equation}
\Gamma^{\prime}(1) = - \gamma 
\label{eq:62}
\end{equation}

\begin{equation}
\Gamma^{\prime}({{3} \over {2}}) = - \sqrt{\pi} \  ({{\gamma} \over {2}} +  \ln 2 - 1) 
\label{eq:63}
\end{equation}

\begin{equation}
\Gamma^{\prime}(2) = - \gamma + 1
\label{eq:64}
\end{equation}

\noindent where $\gamma$ is the Euler's constant. In order to obtain the derivative
of the Hypergeometric Function at issue we resort to the so-called integral
representation, namely \cite{ab}

\begin{equation}
F(s + {{3} \over {2}},1,s + 2, {{3} \over {4}}) = {{\Gamma(s+2)} \over
{\Gamma(1) \  \Gamma(s+1)}} \ \int_0^1 (1 - t)^s \ \left(1 - {{3t} \over {4}} 
\right)^{-s-{{3} \over {2}}} \ dt
\label{eq:65}
\end{equation}

\noindent so that

\begin{equation}
F^{\prime}(s + {{3} \over {2}},1,s + 2, {{3} \over {4}})\vert_{s=0} \  = I_1 +
I_2 + I_3
\label{eq:66}
\end{equation}

\noindent for integrals of the form

\begin{equation}
I_1 = \int_0^1 \left(1 - {{3t} \over {4}}\right)^{-{{3} \over {2}}} \ dt
\label{eq:67}
\end{equation}

\begin{equation}
I_2 = \int_0^1   \left(1 - {{3t} \over {4}}\right)^{-{{3} \over {2}}} 
\ \ln(1 - t) \ dt
\label{eq:68}
\end{equation}

\begin{equation}
I_3 = -  \int_0^1   \left(1 - {{3t} \over {4}}\right)^{-{{3} \over {2}}} \ 
\ln\left(1 - {{3t} \over {4}}\right) \ dt
\label{eq:69}
\end{equation}

If we have that 

\begin{equation}
I_1 = {{8} \over {3}}
\label{eq:70}
\end{equation}

\begin{equation}
I_2 = {{32} \over {3}}  \ln {{2} \over {3}}
\label{eq:71}
\end{equation}

\begin{equation}
I_3 = - {{16} \over {3}} + {{32} \over {3}}  \ln 2
\label{eq:72}
\end{equation}

\noindent then $\zeta_r^{\prime}(0)$ is obviously

\begin{equation}
\zeta_r^{\prime}(0) = 4 \ln 2 + \ln 3
\label{eq:73}
\end{equation}

Under these conditions we can write finally the ratio of determinants $R$ which
appear in (\ref{eq:44}), namely

\begin{equation}
R = {{Det^{\prime} \left[- (d^2/d\tau^2) + V^{\prime \prime}[x_c(\tau)] \right]} \over
{Det \left[- (d^2/d\tau^2) + \omega^2 \right]}}
\label{eq:74}
\end{equation}

In a nutshell

\begin{equation}
R = {{1} \over {12 \omega^2}}
\label{eq:75}
\end{equation}

Combining now the results obtained along this section we write the transition
amplitude anticipated in (\ref{eq:38}), i.e.

\begin{eqnarray*}
<x_f = 1\vert \exp(- H T) \vert x_i = - 1> = 
\end{eqnarray*}
\begin{equation}
 \left({{\omega} \over {\pi}}\right)^{1/2} \
\left(2 \sinh \omega T \right)^{-1/2} \  \sqrt{S_{eo}} \  \sqrt{{{6} \over {\pi}}} \ 
\exp(-S_{eo}) \  \omega \  d\tau_{c} 
\label{eq:76}
\end{equation}

Apart from the first factor, which represents the contribution of the
harmonic oscillator of reference, we have a transition amplitude just
depending on the point $\tau_{c}$ at which the instanton makes the jump.
In accordance with the time interval $T$ the result seems plausible
whenever

\begin{equation}
\sqrt{S_{eo}} \  \sqrt{{{6} \over {\pi}}} \ 
\exp(-S_{eo}) \ \omega \  T \ll 1
\label{eq:77}
\end{equation}

\noindent a nonsense condition if $T$ is large enough. But precisely in this
regime we can accommodate configurations constructed of instantons and
antiinstantons which mimic the behaviour of a trajectory just derived
from the euclidean equation of motion. In doing so we get an additional bonus
since the integration over the parameter $\tau_{c}$ is feasible. As a matter
of fact the level splitting formula is deduced from the complete transition
amplitude once we take into account the multi-instanton configurations. This
serves to exhibit the instanton calculus as an alternative to the WKB approximation
in this context. \par

\section {The dilute-gas approximation.}

Although all the above calculations were carried out over a single instanton,
it remains to identify the hypothetical contributions which incorporate the
effect of a string of instantons and antiinstantons along the $\tau$ axis.
It is customarily assumed that these combinations of topological solutions
represent no strong deviations of the trajectories just derived from the
euclidean equation of motion without any kind of approximation. For background
suppose we have $j$ instantons and antiinstantons centered at points
$\tau_1,...,\tau_j$ whenever

\begin{equation}
-{{T} \over {2}} < \tau_1 < ... < \tau_j < {{T} \over {2}}
\label{eq:78}
\end{equation}

Being narrow enough the regions where the instantons (antiinstantons) make the
jump, the action of the
proposed configuration is almost extremal. We can carry things further and
assume that the action of the string of instantons and antiinstantons is
given by the sum of the $j$ individual actions. This scheme is well-known 
in the literature where appears with the name of dilute gas approximation
\cite{sh}. \par

It should be emphasized at this point that now we can compute amplitudes
with closed paths with $x_i = -1 = x_f$ for instance, so that the action
at issue $S_t$ will be an even multiple of the single instanton action, i.e.

\begin{equation}
S_t \approx 2j \ S_{eo}
\label{eq:79}
\end{equation}

As expected the complete transition amplitude between $x_i = - 1$ and
$x_i =  1$ incorporates the contribution

\begin{equation}
S_t \approx (2j + 1) \ S_{eo}
\label{eq:80}
\end{equation}

In addition the translational degrees of freedom of the  separated $j$
instantons and antiinstantons lead to an integral of the form

\begin{equation}
\int_{-T/2}^{T/2} \omega d\tau_j \
\int_{-T/2}^{\tau_j} \omega d\tau_{j - 1} ... 
\int_{-T/2}^{\tau_2} \omega d\tau_1 = {{(\omega T)^j} \over {j!}} 
\label{eq:81}
\end{equation}

As regards the quadratic fluctuations around the $j$ topological solutions we have
now that the single ratio of determinants transforms into

\begin{eqnarray*}
\left({{\omega} \over {\pi}}\right)^{1/2} \
\left(2 \sinh \omega T \right)^{-1/2} \ 
\left\{{{Det^{\prime} \left[- (d^2/d\tau^2) + V^{\prime \prime}[x_c(\tau)] \right]} \over
{Det \left[- (d^2/d\tau^2) + \omega^2 \right]}}\right\}^{- 1/2} \longrightarrow
\end{eqnarray*}
\begin{equation}
\left({{\omega} \over {\pi}}  \right)^{1/2} \exp(- \omega T/2) \ 
\left[\left\{{{Det^{\prime} \left[- (d^2/d\tau^2) + V^{\prime \prime}[x_c(\tau)] \right]} \over
{Det \left[- (d^2/d\tau^2) + \omega^2 \right]}}\right\}^{- 1/2}\right]^j 
\label{eq:82}
\end{equation}

\noindent according to the limit of the factor associated with the harmonic
oscillator when $T$ is large. With all this information we can write the
complete transition amplitudes for the double-well potential so that

\begin{equation}
<x_f = 1\vert \exp(- H T) \vert x_i = - 1> = 
\left({{\omega} \over {\pi}}  \right)^{1/2} \exp(- \omega T/2)
\sum_{j=1}^{\infty} {{(\omega T d)^{2j+1}} \over {(2j+1)!}}
\label{eq:83}
\end{equation}

\noindent where $d$ stands for the so-called instanton density given by

\begin{equation}
d = \sqrt{{{6} \over {\pi}}} \  \sqrt{S_{eo}} \  \exp(-S_{eo})
\label{eq:84}
\end{equation}

To sum up

\begin{equation}
<x_f = 1\vert \exp(- H T) \vert x_i = - 1> = 
\left({{\omega} \over {\pi}}  \right)^{1/2} \exp(- \omega T/2) \
\sinh (\omega T d)
\label{eq:85}
\end{equation}

Similarly

\begin{equation}
<x_f = 1\vert \exp(- H T) \vert x_i =  1> = 
\left({{\omega} \over {\pi}}  \right)^{1/2} \exp(- \omega T/2)
\sum_{j=0}^{\infty} {{(\omega T d)^{2j}} \over {(2j)!}}
\label{eq:86}
\end{equation}

\noindent so that 

\begin{equation}
<x_f = 1\vert \exp(- H T) \vert x_i =  1> = 
\left({{\omega} \over {\pi}}  \right)^{1/2} \exp(- \omega T/2) \
\cosh (\omega T d)
\label{eq:87}
\end{equation}

Taking the limit $T \rightarrow \infty$ in (\ref{eq:85}), we obtain the
energy eigenvalues $E_0$ and $E_1$ of the first two levels of
the double-well potential

\begin{equation}
E_0 \approx {{\omega} \over {2}} - 2 \omega \sqrt{{{\omega} \over {\pi}}} \exp(-2 \omega/3)
\label{eq:88}
\end{equation}

\begin{equation}
E_1 \approx {{\omega} \over {2}} + 2 \omega \sqrt{{{\omega} \over {\pi}}} \exp(-2 \omega/3)
\label{eq:89}
\end{equation}

As anticipated in Section III the quantum mechanical tunneling transfers the
wave function from one well to the other, thus lifting the degeneracy of the
classical vacua. To close, notice the way in which the energy eigenvalues
depend on the barrier-penetration factor, i.e. the exponential of  minus
the classical action of the instanton. \par

\vfill \eject

\section {Appendix A.}

In this appendix we include some results concerning the shape-invariance
symmetry in one-dimensional 
supersymmetric quantum mechaniccs (susy qm for short). First of all one
can assume the existence of two first-order differential operators
$A$, $A^{\dagger}$ which in terms of the superpotential function $W(z)$
are of the form

\begin{equation}
A = {{d} \over {dz}} + W(z)
\label{eq:201}
\end{equation}
 
\begin{equation}
A^{\dagger} = - {{d} \over {dz}} + W(z) 
\label{eq:202}
\end{equation}

This enables us to write now a couple of hamiltonians $H_{-}$ and $H_{+}$ given by
 
\begin{equation}
H_{-} = A^{\dagger} A 
\label{eq:203}
\end{equation}

\begin{equation}
H_{+} = A A^{\dagger}
\label{eq:204}
\end{equation}
 
To be precise

\begin{equation}
H_{\pm} = - {{d^2} \over {dz^2}} + V_{\pm}(z) 
\label{eq:205}
\end{equation}

\noindent for 

\begin{equation}
V_{\pm}(z) = W^2(z) \pm W'(z) 
\label{eq:206}
\end{equation}

\noindent where the prime denotes as usual the
derivative with respect to the spatial coordinate.
It is often the case to refer to the pair of potentials
$V_{-}(z)$ and $V_{+}(z)$ as supersymmetric partners in accordance with the
standard susy qm language. 
Now we look for the supersymmetric form of a standard quantum mechanical model
associated with the hamiltonian $H_{V}$ given by

\begin{equation}
H_{V}= - {{d^2} \over {dz^2}} + V(z)
\label{eq:207}
\end{equation}

\noindent where $V(z)$ is some well-behaved function such that $H_{V}$ has a
non-empty discrete spectrum which is below a possible continuous spectrum. It
seems plausible to identify $H_{V}$ with the hamiltonian $H_{-}$ of a standard
susy qm model. Being $\epsilon$ the energy of the ground state $\phi_{o}(z)$
of $H_{V}$ we can write that 

\begin{equation}
H_{V}= H_{-} + \epsilon
\label{eq:208}
\end{equation}

\noindent so that $\epsilon$ represents the role of the factorization energy
while the superpotential $W(z)$ which appears in the first differential
operators $A$ and $A^{\dagger}$ is given by \cite{ju}

\begin{equation}
W(z) = - {{\phi_{o}^{\prime}(z)} \over {\phi_{o}(z)}} 
\label{eq:209}
\end{equation}

In other words, we are dealing with a model where the ground
state of $H_{V}$ is nothing but the zero-mode of $H_{-}$.
Once written $H_{+}$ in the usual way, the two partner hamiltonians $H_{-}$ and
$H_{+}$ are essential isospectral: the ground state energy of $H_{-}$ vanishes
and all other eigenvalues coincide with that of $H_{+}$. In such a case the
hamiltonian $\tilde{H}_{V}$, defined as

\begin{equation}
\tilde{H}_{V} = H_{+} + \epsilon
\label{eq:210}
\end{equation}

\noindent exhibits a spectrum which coincides with the set of energy eigenvalues
of $H_{V}$ except for $\epsilon$ itself. It is obvious that by repetition of
this method we can find a hierarchy of essential isospectral hamiltonians. \par

At this point we can expose in brief how the so-called shape-invariance property
allows to solve in a closed form the Schrodinger's equation. Let us assume the
existence of a  susy qm system described in terms
of a superpotential $W(z,a_{o})$ which depends on the parameter $a_{o}$. The
partner potentials $V_{\pm}(z,a_{o})$ given by

\begin{equation}
V_{\pm}(z,a_{o}) = W^{2}(z,a_{o}) \pm W^{\prime}(z,a_{o})
\label{eq:211}
\end{equation}

\noindent are called shape-invariant if they are related by 

\begin{equation}
V_{+}(z,a_{o}) = V_{-}(z,a_{1}) + R(a_{1})
\label{eq:212}
\end{equation}

\noindent where $a_{1}$ is a new parameter written in terms of $a_{o}$ as
$a_{1} = F(a_{o})$ while the residual contribution $R(a_{1})$ is in fact
independent of the variable $z$. To sum up, the shape-invariance property
means that the partner potential $V_{+}(z,a_{o})$ can be considered as a new
potential $V_{-}(z,a_{1})$ with superpotential $W(z,a_{1})$ once the $R(a_{1})$
has been subtracted. 
If the mapping $a_{s} = F(a_{s-1})$ may be
iterated  leading to a family of well-behaved superpotentials $W(z,a_{s})$, then
the energy eigenvalues and their corresponding wave functions can be obtained
in a simple way \cite{ju}. \par
As an example we focus our interest on the set of
hamiltonians $O_{\ell}$ given by

\begin{equation}
O_{\ell} = - {{d^2} \over {dz^2}} - {{\ell(\ell+1)} \over {\cosh^{2} z}} + \ell^2
\label{eq:213}
\end{equation}

\noindent where $\ell =1, 2, ...$, although only the first members of the series
are relevant in physics. It is the case that $O_{\ell}$ can be factorized in terms
of a superpotential $W(z,\ell)$ like

\begin{equation}
W(z,\ell) = \ell \tanh z
\label{eq:214}
\end{equation}

\noindent so that $O_{\ell} = A_{\ell}^{\dagger}A_{\ell}$ (notice that the
energy of factorization $\epsilon$ vanishes) for

\begin{equation}
A_{\ell} = {{d} \over {dz}} + \ell \tanh z
\label{eq:215}
\end{equation}

\begin{equation}
A_{\ell}^{\dagger} = - {{d} \over {dz}} + \ell \tanh z
\label{eq:216}
\end{equation}

The shape-invariance condition appears once we write the partner hamiltonian
$\tilde{O}_{\ell}$ so that in this example the mapping between the old
parameter $\ell$ and the new one $\tilde{\ell}$ reduces to

\begin{equation}
\tilde{\ell} = \ell - 1
\label{eq:217}
\end{equation}

The procedure may be iterated leading to a family of well-behaved superpotentials 
which allow us to extract the physical content of $O_{\ell}$. The discrete spectrum
includes a normalizable zero-energy mode $\phi_{\ell,o}(z)$ of the form

\begin{equation}
\phi_{\ell,o}(z) = {{\sqrt{2(2\ell - 1)!}} \over {2^{\ell} (\ell - 1)!}} \
{{1} \over {\cosh^{\ell} z}}
\label{eq:218}
\end{equation}

\noindent which is below the set of states $\phi_{\ell,m}(z)$ given by

\begin{equation}
\phi_{\ell,m}(z) = {{\sqrt{2(2\ell - 2 m - 1)!}} \over {2^{\ell - m} 
(\ell - m -  1)!}} \ {{1} \over {\sqrt{\prod _{j = 0}^{m - 1} (E_{m} - E_{j})}}} 
\ A_{\ell}^{\dagger}(z) \ .\ .\ . \
A_{\ell - m + 1}^{\dagger}(z) \ \left[{{1} \over {\cosh^{\ell - m} z}}\right]
\label{eq:219}
\end{equation}

\noindent with energies

\begin{equation}
E_{m} = \ell^{2} - (\ell - m)^2
\label{eq:220}
\end{equation}

\noindent for $m = 1, ... , \ell - 1$. The continuous spectrum $\phi_{\ell,k}(z)$
corresponds to

\begin{equation}
\phi_{\ell,k}(z) = {{A_{\ell}^{\dagger}(z)} \over {\sqrt{k^2 + \ell^2}}} \
{{A_{\ell - 1}^{\dagger}(z)} \over {\sqrt{k^2 + (\ell - 1)^2}}} \ .\ .\ . \
{{A_{1}^{\dagger}(z)} \over {\sqrt{k^2 + 1}}} \ \left[{{\exp(ikz)} \over
{\sqrt{2 \pi}}} \right]
\label{eq:221}
\end{equation}

\noindent with energy eigenvalues

\begin{equation}
E_{k} = k^2 + \ell^2
\label{eq:222}
\end{equation}

\noindent and standard normalization as follows

\begin{equation}
\int_{- \infty}^{\infty} \ \phi_{\ell,k}^{*}(z) \ \phi_{\ell,k^{\prime}}(z) \ dz = \delta(k - k^{\prime})
\label{eq:223}
\end{equation}

\vfill \eject

\section {Appendix B.}

The first quantum correction to the classical term in the path-integral formalism
is calculated by evaluating the determinant of a differential operator. As the
determinant includes the product of the eigenvalues at issue the result would be
in principle a terribly divergent expression. As far as we know Hawking \cite{ha}
was the first to consider in a systematic way the zeta-function method as a useful
technique for the study of these determinants. For
reference, let us  take a one-dimensional hamiltonian $H$ with positive
discrete eigenvalues $a_j$. If the corresponding eigenfunctions are $\phi_j(z)$ 
one can write that

\begin{equation}
H\phi_j(z) = a_j \phi_j(z)
\label{eq:224}
\end{equation}

Next we introduce the so-called zeta-function associated with $H$, i.e.

\begin{equation}
\zeta_{H}(s) = \sum_{j} {{1} \over {a_j^s}}
\label{eq:225}
\end{equation}

Now if one notes that

\begin{equation}
{{d\zeta_{H}(s)} \over {ds}} = - \sum_j \ln a_j \ \exp(-s\ln a_j)
\label{eq:226}
\end{equation}

\noindent the determinant of $H$ is given by

\begin{equation}
Det \  H = \exp-\zeta_{H}^{\prime}(0)
\label{eq:227}
\end{equation}

This structure is enriched by the existence of the heat-kernel function
$G(z,w,\mu)$, i.e.

\begin{equation}
G(z,w,\mu) = \sum_j \exp(-a_j \mu) \ \ \phi_j^{*}(z) \phi_j(w)
\label{eq:228}
\end{equation}

\noindent which satisfies the heat difussion equation

\begin{equation}
H_z G(z,w,\mu) = - {{\partial G(z,w,\mu)} \over {\partial \mu}}
\label{eq:229}
\end{equation}

\noindent with an initial condition like

\begin{equation}
G(z,w,0) = \delta(z - w)
\label{eq:230}
\end{equation}

On the other hand $\zeta_{H}(s)$ can be understood as the Mellin transformation
of $G(z,w,\tau)$ so that

\begin{equation}
\zeta_{H}(s) = {{1} \over {\Gamma(s)}} \int_0^{\infty} \mu^{s-1} \ d\mu \ 
\int_{-\infty}^{\infty} \ G(z,z,\mu) \ dz
\label{eq:231}
\end{equation}

To sum up, if we know the eigenfunctions $\phi_j(z)$ and the eigenvalues $a_j$
of the hamiltonian $H$ it suffices to insert such information in (\ref{eq:228})
to compute
$\zeta_{H}(s)$. Finally we use (\ref{eq:227}) to obtain $Det \ H$. This
zeta-function method makes it simple to study the scaling of the functional
determinants we are involved with. Under a transformation of the form

\begin{equation}
H \rightarrow \tilde{H} = \beta H
\label{eq:232}
\end{equation}

\noindent where $\beta$ represents a constant factor, one finds that

\begin{equation}
Det \  \tilde{H} = \beta^{\zeta_{H}(0)} Det \ H
\label{eq:233}
\end{equation}

\noindent as can be seen by inspection. However, this is not yet the end of
the story since we need to adapt this method to the general case where the
operator $H$ also includes a continuous spectrum. To fix the ideas let us now
sketch the way in which one can calculate the ratio of the determinants associated
with $O_{\ell}$ (see appendix A) and $P_{\ell}$ which is given by

\begin{equation}
P_{\ell} = - {{d^2} \over {dz^2}} + \ell^2
\label{eq:234}
\end{equation}

Notice that $P_{\ell}$ itself represents a free-particle comparison hamiltonian for 
$O_{\ell}$. To sum up we need to study

\begin{equation}
Q_{\ell} = {{Det^{\prime} \ O_{\ell}} \over {Det \ P_{\ell}}}
\label{eq:235}
\end{equation}

\noindent where $Det^{\prime} \ O_{\ell}$ denotes the reduced determinant
obtained once the zero-mode has been explicitly removed. To handle the
continuous spectrum we resort to the density matrix $\Xi_{O_{\ell}}(k,z,w)$
written as \cite{ni}

\begin{equation}
\Xi_{O_{\ell}}(k,z,w) = \phi_{\ell,k}^{*}(z) \ \phi_{\ell,k}(w)
\label{eq:236}
\end{equation}

When going to $P_{\ell}$ we get

\begin{equation}
\Upsilon_{P_{\ell}}(k,z,w) = {{\exp\left[-ik(z - w)\right]} \over {2 \pi}}
\label{eq:237}
\end{equation}

\noindent as corresponds to a free-particle model. An additional 
subtraction prescription leads us to the regularized spectral density 
$\rho_r(k)$ expressed as

\begin{equation}
\rho_r(k) = \int_{-\infty}^{\infty} \ \left[\Xi_{O_{\ell}}(k,z,z) - 
\Upsilon_{P_{\ell}}(k,z,z)\right] \ dz
\label{eq:238}
\end{equation}

In this framework the appropiate zeta-function for the analysis of (\ref{eq:235})
would be

\begin{equation}
\zeta_r(s) = \zeta_{O_{\ell}}(s) - \zeta_{P_{\ell}}(s)
\label{eq:239}
\end{equation}

\noindent so that

\begin{equation}
\zeta_r(s) = {{1} \over {\Gamma(s)}} \int_0^{\infty} \mu^{s-1}  d\mu  
\left\{\exp\left[- \sum_{m = 1}^{\ell - 1} [\ell^2 - (\ell - m)^2] \mu \right] +
\int_{-\infty}^{\infty}  \rho_r(k) \exp[-(k^2 + \ell^2) \mu] \  dk \right\}
\label{eq:240}
\end{equation}

\vfill \eject


\begin{thebibliography}{99}

\bibitem{po}
{A. Polyakov,} Nucl. Phys. {\bf B120} (1977) 429.
\bibitem{ha}
{S.W. Hawking,} Commun. Math. Phys. {\bf 55} (1977) 133.
\bibitem{kl}
{H. Kleinert,} {\it Paths Integrals in Quantum Mechanics, Statistics and
Polymer Physics}.
Singapore: World Scientific (1990).
\bibitem{ra}
{P. Ramond,} {\it Field Theory: A Modern Primer}.
New York: Addison-Wesley Publishing Company (1992). 
\bibitem{gr}
{I.S. Gradshteyn and I.M. Ryzhik,} {\it Table of Integrals, Series and Products}.
New York: Academic Press (1965). 
\bibitem{ju}
{G. Junker,} {\it Supersymmetric Methods in Quantum and Statistical Physics}.
Berlin: Springer-Verlag (1996).
\bibitem{ab}
{M. Abramowitz and I.A. Stegun,} {\it Handbook of Mathematical Functions}.
New York: Dover Pu\-bli\-ca\-tions (1970).
\bibitem{sh}
{M.A. Shifman,} {\it ITEP Lectures on Particle Physics and Field Theory}.
Singapore: World Scientific (1999).
\bibitem{ni}
{A.J. Niemi and G.W. Semenoff,}  Phys. Rep. {\bf 135} (1986) 99.



\end{thebibliography}
\end{document}